\documentclass{aa}  

\usepackage{color}

\usepackage{multirow}
\usepackage{changes}
\usepackage{orcidlink}
\usepackage[switch]{lineno} 
\defcitealias{Nicolas_2021}{N21} 

\usepackage{graphicx}
\usepackage{txfonts}
\hypersetup{
    colorlinks=true,
    linkcolor=blue,
    filecolor=blue,      
    urlcolor=blue,
    citecolor=blue
    }

\begin{document} 

   \title{ZTF SN Ia DR2: Colour standardisation of type Ia supernovae and its dependence on the environment}

   \author{
        Ginolin, M. \inst{1} \fnmsep\thanks{Corresponding author: \texttt{m.ginolin@ip2i.in2p3.fr}} \orcidlink{0009-0004-5311-9301}
        \and Rigault, M. \inst{1} \orcidlink{0000-0002-8121-2560}
        \and Copin, Y. \inst{1} \orcidlink{0000-0002-5317-7518}
        \and Popovic, B. \inst{1} \orcidlink{0000-0002-8012-6978}
        \and Dimitriadis, G. \inst{2} \orcidlink{0000-0001-9494-179X}
        \and Goobar, A. \inst{3} \orcidlink{0000-0002-4163-4996}
        \and Johansson, J. \inst{3} \orcidlink{0000-0001-5975-290X}
        \and Maguire, K.\inst{2} \orcidlink{0000-0002-9770-3508}
        \and Nordin, J. \inst{4} \orcidlink{0000-0001-8342-6274}
        \and Smith, M. \inst{5} \orcidlink{0000-0002-3321-1432}
        \and Aubert, M. \inst{6}
        \and Barjou-Delayre, C. \inst{6}
        \and Burgaz, U.\inst{2} \orcidlink{0000-0003-0126-3999}
        \and Carreres, B. \inst{7, 8} \orcidlink{0000-0002-7234-844X}
        \and Dhawan, S. \inst{9} \orcidlink{0000-0002-2376-6979}
        \and Deckers, M. \inst{2} \orcidlink{0000-0001-8857-9843}
        \and Feinstein, F. \inst{7}
        \and Fouchez, D. \inst{7} \orcidlink{0000-0002-7496-3796}
        \and Galbany, L. \inst{10, 11} \orcidlink{0000-0002-1296-6887}
        \and Ganot, C. \inst{1}
        \and de Jaeger, T. \inst{12} \orcidlink{0000-0001-6069-1139}
        \and Kim, Y.-L. \inst{5} \orcidlink{0000-0002-1031-0796}
        \and Kuhn, D. \inst{12} \orcidlink{0009-0005-8110-397X}
        \and Lacroix, L. \inst{3, 12} \orcidlink{0000-0003-0629-5746}
        \and M\"uller-Bravo, T. E. \inst{10, 11} \orcidlink{0000-0003-3939-7167}
        \and Nugent, P. \inst{13, 14} \orcidlink{0000-0002-3389-0586}
        \and Racine, B. \inst{7} \orcidlink{0000-0001-8861-3052}
        \and Rosnet, P. \inst{6} \orcidlink{0000-0002-6099-7565}  
        \and Rosselli, D. \inst{7} \orcidlink{0000-0001-6839-1421}
        \and Ruppin, F. \inst{1} \orcidlink{0000-0002-0955-8954}
        \and Sollerman, J. \inst{15} \orcidlink{0000-0003-1546-6615}
        \and Terwel, J.H. \inst{2, 16} \orcidlink{0000-0001-9834-3439}
        \and Townsend, A. \inst{4} \orcidlink{0000-0001-6343-3362}
        \and Dekany, R. \inst{17} \orcidlink{0000-0002-5884-7867}
        \and Graham, M. \inst{18} \orcidlink{0000-0002-3168-0139}
        \and Kasliwal, M. \inst{18} \orcidlink{0000-0002-5619-4938}
        \and Groom, S.L. \inst{19} \orcidlink{0000-0001-5668-3507}
        \and Purdum, J. \inst{17} \orcidlink{0000-0003-1227-3738}
        \and Rusholme, B. \inst{19} \orcidlink{0000-0001-7648-4142}
        \and van der Walt, S. \inst{20} \orcidlink{0000-0001-9276-1891}
          }

    \institute{Univ Lyon, Univ Claude Bernard Lyon 1, CNRS, IP2I Lyon/IN2P3, UMR 5822, F-69622, Villeurbanne, France
    \and School of Physics, Trinity College Dublin, College Green, Dublin 2, Ireland
    \and Oskar Klein Centre, Department of Physics, Stockholm University, SE-10691 Stockholm, Sweden
    \and Institut für Physik, Humboldt Universität zu Berlin, Newtonstr 15, 12101 Berlin, Germany
    \and Department of Physics, Lancaster University, Lancs LA1 4YB, UK
    \and Université Clermont Auvergne, CNRS/IN2P3, LPCA, F-63000 Clermont-Ferrand, France
    \and Aix Marseille Université, CNRS/IN2P3, CPPM, Marseille, France
    \and Department of Physics, Duke University, Durham, NC 27708, USA
    \and Institute of Astronomy and Kavli Institute for Cosmology, University of Cambridge, Madingley Road, Cambridge CB3 0HA, UK
    \and Institute of Space Sciences (ICE, CSIC), Campus UAB, Carrer de Can Magrans, s/n, E-08193, Barcelona, Spain
    \and Institut d'Estudis Espacials de Catalunya (IEEC), E-08034 Barcelona, Spain
    \and Sorbonne Université, CNRS/IN2P3, LPNHE, F-75005, Paris, France
    \and Lawrence Berkeley National Laboratory, 1 Cyclotron Road MS 50B-4206, Berkeley, CA, 94720, USA
    \and Department of Astronomy, University of California, Berkeley, 501 Campbell Hall, Berkeley, CA 94720, USA
    \and Oskar Klein Centre, Department of Astronomy, Stockholm University, SE-10691 Stockholm, Sweden
    \and Nordic Optical Telescope, Rambla José Ana Fernández Pérez 7, ES-38711 Breña Baja, Spain
    \and Caltech Optical Observatories, California Institute of Technology, Pasadena, CA 91125
    \and Division of Physics, Mathematics, and Astronomy, California Institute of Technology, Pasadena, CA 91125, USA
    \and IPAC, California Institute of Technology, 1200 E. California Blvd, Pasadena, CA 91125, USA
    \and Berkeley Institute for Data Science, University of California Berkeley, Berkeley, CA 94720, USA
    }

   \date{Received;}
 
  \abstract
   {As type Ia supernova cosmology transitions from a statistics-dominated to a systematics-dominated era, it is crucial to understand the remaining unexplained uncertainties that affect their luminosity, such as those stemming from astrophysical biases. Type Ia supernovae are standardisable candles whose absolute magnitude reaches a scatter of typically 0.15~mag when empirical correlations with their light-curve stretch and colour and with their environmental properties are accounted for.}
   {We investigate the dependence of the standardisation process of type Ia supernovae on the astrophysical environment to ultimately reduce their scatter in magnitude. We focus on colour standardisation.}
   {We used the volume-limited ZTF SN Ia DR2 sample, which offers unprecedented statistics for the low-redshift ($z<0.06$) range. We first studied the colour distribution with a focus on the effects of dust to then select a dustless subsample of objects that originated in environments with a low stellar mass and in the outskirts of their host galaxies. We then examined the colour-residual relation and its associated parameter $\beta$. Finally, we investigated the colour dependence of the environment-dependent magnitude offsets (steps) to separate their intrinsic and extrinsic components.}
   {Our sample of nearly 1,000 supernovae probes the red tail of the colour distribution up to $c=0.8$. The dustless sample exhibits a significantly shorter red tail ($4.3\sigma$) than the whole sample, but the distributions around $c\sim0$ are similar for both samples. This suggests that the reddening above $c\geq0.2$ is dominated by interstellar dust absorption of the host and that the remaining colour scatter has an intrinsic origin. The colour-residual relation is linear with light-curve colour. We found indications of a potential evolution of $\beta$ with the stellar host mass, with $\beta\sim3.6$ for low-mass galaxies, compared to $\beta=3.05\pm0.06$ for the full sample. Finally, in contrast to recent claims from the literature, we found no evolution of steps as a function of light-curve colour. This suggests that dust may not be the dominating mechanism for the dependence on the environment of the magnitude of type Ia supernovae.}
   {}


    \titlerunning{ZTF SN Ia DR2: Colour standardisation of type Ia supernovae and its dependence on the environment}
    \authorrunning{Ginolin, M.}
    
   \maketitle
%

\section{Introduction}
\label{sec:intro}

Type Ia supernovae (SNe Ia) were the tools that enabled the discovery of the accelerated expansion of the Universe because they can be standardised \citep{Riess_1998, Perlumtter_1999}. They are still a crucial probe for precision cosmology, especially in the determination of the dark energy equation-of-state parameter $w$ \citep{Planck_2020, Brout_2022} and of the Hubble-Lemaître constant $H_0$ \citep{Freedman_2021, Riess_2022} because they are a unique tracer of the expansion of the local Universe when they are combined with a distance calibrator (e.g. Cepheids).

The scatter of the raw observed magnitudes of SNe Ia is $\sim0.4$ mag. Part of this scatter can be reduced through the empirical relations discovered in the mid-1990s between the SN magnitudes and their colour and between the SN magnitudes and their stretch \citep{Phillips_1993, Tripp_1998}. Based on these relations, the scatter can be decreased to $\sim0.15$ mag, and the residual scatter is dominated by a scatter that is called intrinsic, for lack of better understanding. 
In the mid-2010s, an additional standardisation relation between the SN host mass and their magnitude was brought to light \citep{Kelly_2010, Sullivan_2010, Lampeitl_2010, Childress_2013}. This accounts for astrophysical biases on the SN Ia luminosity. In contrast to the stretch and colour-magnitude relations, which are linear, this correction was typically parametrised as a magnitude offset between high- and low-mass hosts. It is called the 'mass step'.

While they are widely used in cosmology, the formation mechanism of SNe Ia is still unclear. Type Ia SNe are thought to be the thermonuclear explosion of a carbon-oxygen white dwarf. However, several scenarios coexist in the literature (see \citealt{Maeda_Terada_2016} for a review): For example, the single-degenerate scenario, according to which a white dwarf accretes enough material from a companion star to reach masses that are high enough to reach instability regimes, or the double-degenerate scenario, according to which an explosion occurs after a merger of two white dwarfs or the accretion of one white dwarf by the other. 
The lack of an understanding of this mechanism, both theoretically and through first-principle numerical simulations, also means that the link between the SN~Ia properties and their environments is largely unknown. The stellar mass, stellar age, gas and stellar metallicities, and/or the star formation rate of the host might play various roles in the observational characteristics of SN Ia, but it is unclear to which extent this is the case.

In this context, the origin of the SN~Ia colour, whose correlation with magnitude provides the strongest reduction in scatter in the standardisation process, is often discussed. Current interpretations suggest that it is a mixture of the intrinsic colour that results from the poorly understood formation mechanisms and additional reddening by interstellar dust of the host on the line of sight \citep[e.g.][]{jha_2007, Mandel_2017, Uddin_2020, Brout_Scolnic_2021, Johansson_2021, Kelsey_2023, Duarte_2023, Grayling_2024}. The typical SN~Ia standardisation still assumes a unique  linear colour-magnitude relation coefficient $\beta$, which might not describe the data correctly if indeed the colour origin is twofold. Furthermore, this single $\beta$ coefficient is found to be significantly lower than what would be expected from nearby galaxy dust models and data \citep{Wang_2005, Nobili_Goobar_2008, Goobar_2008, Amanullah_2015, Salim_2018}. 

In parallel, the stretch- and colour-standardised SN~Ia magnitudes depend significantly on the SN host properties, such that SNe Ia from massive, red, and/or passive galaxies are significantly brighter than those from low-mass, blue, and/or star-forming environments \citep[e.g][]{Sullivan_2010, Childress_2013, Rigault_2013, Kim_2018, Rigault_2020, Kelsey_2021}. The exact origin of these so-called magnitude steps is highly debated. For historical reasons and simplicity, however, the mass step is commonly used in current cosmological analyses \citep{Betoule_2014, Scolnic_2018, Brout_2022, Riess_2022}. Two explanations are currently in the focus of the discussions. The first assumes an intrinsic origin of the environmental magnitude offset that is related to the progenitor age, in the sense of a prompt versus delayed dichotomy \citep[e.g.][]{Sullivan_2006}, so that prompt SNe~Ia are intrinsically fainter after stretch and colour standardisation than delayed SN Ia \citep{Rigault_2013, Rigault_2020, Briday_2021}. Alternatively, interstellar dust may vary as a function of the environment, such that massive hosts have both more dust and different dust properties than low-mass hosts \citep[][see also \citealt{Gonzalez-Gaitan_2021, Wojtak_2023}]{Brout_Scolnic_2021, Popovic_2021, Meldorf_2023}. The most recent cosmological results \citep{Brout_2022, Riess_2022, Vincenzi_2024} assumed the latter \citep[see][and references therein]{Popovic_2023}. In practice, both effects are likely to play a role, and the question then is the relative amplitude of each \citep{Wiseman_2022, Kelsey_2023}.

An avenue to solve this question is to investigate the environmental dependence of SNe~Ia that are observed in the near-infrared (NIR), where the effect of dust is expected to be negligible \citep{Jones_2022_RAISIN, Galbany_2023}, and where the SN~Ia magnitudes are naturally less scattered \citep[e.g.][]{Barone-Nugent_2012, Dhawan_2018}. If the environmental magnitude offsets were solely due to varying dust properties, the NIR data would not be expected to exhibit any mass step. \cite{Uddin_2020, Uddin_2023}, and \cite{Ponder_2021} reported significant steps, however, with similar amplitudes as were found using optical data. This strongly disfavours the dust scenario. However, \cite{Johansson_2021} and \cite{Jones_2022_RAISIN} reported more mitigated results. They also found steps, but not as significant ($\sim2\sigma$ level). This indeed suggests that NIR SN~Ia data might not be affected the environmental biases described above.

We investigate the colour part of SN~Ia standardisation in detail based on the second data release of the Zwicky Transient Facility \citep[ZTF,][]{Bellm_2019, Graham_2019, Masci_2019, Dekany_2020} cosmology group (SN Ia DR2, \citealt{DR2_overview, Smith_2024}). 
A companion paper \citep{Ginolin_2024a} focused on stretch standardisation and on the typical environment-dependent magnitude offsets (steps). In this paper, we showed based on a $\mathcal{O}(1000)$ volume-limited SN~Ia dataset that the stretch distribution follows what was expected by the so-called prompt and delayed model from \cite{Rigault_2020} and \cite{Nicolas_2021}, and that the ZTF SNe Ia exhibit a strong environmental bias ($\geq 0.15~\mathrm{mag}$). In this paper, we investigate the origin of this step in the context of the discussion of age versus dust.

After a brief introduction of the ZTF DR2 data in Section~\ref{sec:data}, we first study the ZTF SN~Ia colour distribution and its dependence on the SN environment in Section \ref{sec:colour_distribution}. In Section \ref{sec:colour_mag_relation}, we examine the dependence of the colour-standardisation parameter $\beta$ on the SN hosts. In this section, we discuss the linearity of the colour-magnitude relation in the context of the twofold origin of the SN colour. The origin of the SN~Ia magnitude astrophysical dependence is studied in Section \ref{sec:steps}, where we analyse the variation in the magnitude offsets as a function of SN colour. We finally discuss the robustness of our results and their consequences for SN Ia cosmology in Section~\ref{sec:discussion}. We conclude in Section~\ref{sec:conclusion}.

\section{Data}
\label{sec:data}

\subsection{Zwicky Transient Facility SN Ia DR2}
\label{sec:ztfdr2}

We used the ZTF SN Ia DR2 sample described in \cite{DR2_overview}. The ZTF DR2 data were presented in detail in \cite{Smith_2024}. In this analysis, we probe the properties of the underlying SNe~Ia population, and we therefore limited our sample to well-measured objects in the volume-limited redshift range ($z<0.06$). 

As described in \cite{DR2_overview}, these well-measured SNe~Ia meet the following two quality cuts: (1) At least seven $5\sigma$ detections, two pre- and two-post maximum light and in two different filters in the rest-frame phase range [$-10$ days, $+40$ days] that is used to fit the \texttt{SALT2} parameters \citep{Guy_2010,Betoule_2014, Taylor_2021}. (2) A \texttt{SALT2} light-curve fit probability greater than $10^{-7}$. 
We fit the ZTF light curves with the \texttt{SALT2} T21 model \citep{Taylor_2021}, and we account for Milky Way extinction (see details in \cite{Smith_2024}). We limited our targets to those with the following \texttt{SALT2} parameters: $c\in [-0.2, 0.8]$, with $\sigma_c<0.1$, $x_1 \in [-3, +3]$, with $\sigma_{x_1}<1$, and a peak-luminosity date $t_0$ measured with a precision of one day or less. This left 993 SNe~Ia. As recommended by \cite{rose2022}, we extended the colour range above the usual $c<0.3$, which allowed us to probe the characteristics of the full SN~Ia colour distribution. In addition, we removed $\sim20$ peculiar SNe Ia (e.g. 91bg or Ia-CSM), but we kept edge cases such as 91T because they typically cannot be identified in higher-redshift samples (see details in \citealt{Dimitriadis_2024}). We show in Sect.~\ref{sec:tests} that our results do not depend on these cuts. We also removed $\sim 30$ additional objects that lacked measurements of the host properties, which left 945 SNe~Ia. Finally, to be consistent throughout the analysis, we removed an additional 7 objects that were discarded by the outlier rejection in the Hubble residual standardisation (see Sect. \ref{sec:colour_mag_relation}). 

Our final sample therefore contained 938 SNe. In this sample, the redshift of 76\% SNe comes from the spectroscopic features of their hosts, mostly from the MOST Hosts DESI program \citep{MOST_Hosts}, with a typical precision greater that $\sigma_z\leq10^{-4}$. The redshift of the remaining 24\% comes from the spectroscopic features of the SN. As detailed in \cite{Smith_2024}, these redshifts are not biased and have a typical precision of $\sigma_z\leq3\times10^{-3}$.

\subsection{Environmental cuts}
\label{sec:envcuts}

To study the environmental dependence of SNe Ia, we used four optical photometric tracers made available as part of the SN Ia DR2: the local (2 kpc radius) and global \texttt{PS1.g}-\texttt{PS1.z} colour \citep[PS1: Pan-STARRS 1, ][]{Pan-STARRS} and local and global stellar mass, computed with the \texttt{HostPhot} package (\citealt{HOSTPHOT}; see details in \citealt{Smith_2024}). The colours and photometric masses were corrected for Milky Way extinction. When SNe are compared according to their environments, the SNe~Ia are split into two subsamples using one of the following cuts:
\begin{itemize}
    \item $(g-z)_\mathrm{local, global}^\mathrm{cut}=1$, which corresponds to the gap in the colour distribution.
    \item $\log(M_\star/M_\odot)_\mathrm{global}^\mathrm{cut}=10$, as is commonly done in the literature.
    \item $\log(M_\star/M_\odot)_\mathrm{local}^\mathrm{cut}=8.9$, the median of the local mass distribution.
\end{itemize}

\section{Colour distribution}
\label{sec:colour_distribution}

\begin{figure}
    \centering
    \includegraphics[width=1\columnwidth]{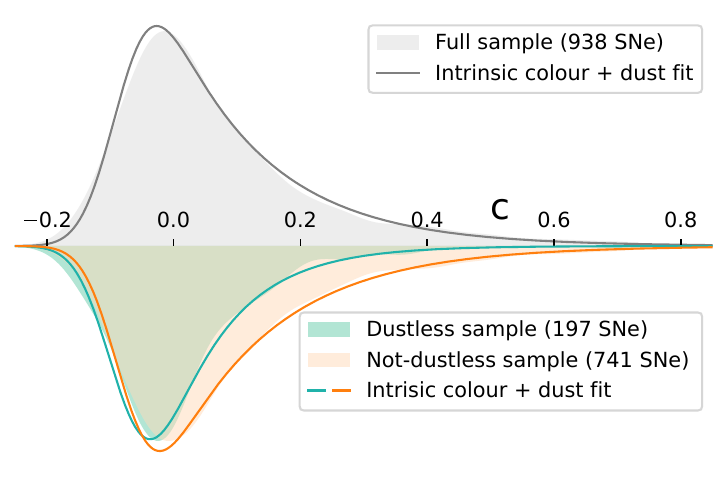}
    \caption{Ideogram of SN colour and the fit described in Eq. \ref{eq:dust} for the full sample (top plot), the dustless, and the not-dustless samples (bottom plot). The mean error on $c$ is added in quadrature to $\sigma_c$ in the plotted fits because we took the errors on $c$ into account in our model.}
    \label{fig:colour_distrib}
\end{figure}

In this section, we investigate the SN colour ($c$) distribution. The \texttt{SALT2} \citep[“T21”,][]{Taylor_2021} colour distribution of our sample is shown in Fig.~\ref{fig:colour_distrib}. A clear asymmetry is visible in this distribution, with a tail extending redward, as pointed out by \cite{Brout_Scolnic_2021} for a different compilation of SNe Ia.

Following former analyses \cite[e.g.][]{jha_2007, Mandel_2011, Mandel_2017}, we modelled the colour distribution as the convolution of a normal distribution $\mathcal{N}(c\,|\,c_\mathrm{int}, \sigma_c)$ with an exponential decay for positive colour, such that
\begin{equation}
    P(c)= \mathcal{N}(c\,|\,c_\mathrm{int}, \sigma_c) \otimes \begin{cases} 0 & \mbox{if } c \leq 0 \\ \frac{1}{\tau}e^{-c/\tau} & \mbox{if } c > 0.
    \end{cases}
    \label{eq:dust}
\end{equation}
Conceptually, the normal distribution should represent the intrinsic SN~Ia colour scatter, while the exponential decay accounts for additional reddening due to the interstellar dust of the host galaxy. We evaluated $P(c_i)$ from Eq. \ref{eq:dust} for the ith SN and then minimised $\sum_i-\log(P(c_i))$ with \texttt{iminuit} \citep{iminuit}. The measurement errors on $c$ are added in quadrature with $\sigma_c$. 

The best-fit model for the ZTF data is shown in Fig.~\ref{fig:colour_distrib}, and the values of the parameters are presented in Table \ref{tab:colour_model}. It reproduces the data qualitatively well. This supports the idea of an intrinsic and extrinsic origin of the observed SN Ia colours.

\begin{table}
\centering
\tiny
\caption{Best-fit values for the parameters of Eq. \ref{eq:dust}.}
\label{tab:colour_model}
\begin{tabular}{l  c  c  c } 
\hline\\[-0.8em]
\hline\\[-0.5em]
Sample & $c_\mathrm{int}$ & $\sigma_c$ & $\tau$\\
\hline\\[-0.5em]
Full sample & $-0.085\pm0.004$ & $0.030\pm0.005$ & $0.155\pm0.007$ \\[0.30em]
Dustless sample & $-0.087\pm0.008$ & $0.033\pm0.008$ & $0.102\pm0.010$ \\[0.30em]
Non dustless sample & $-0.082\pm0.005$ & $0.029\pm0.006$ & $0.166\pm0.008$  \\[0.30em]
\hline
\end{tabular}
\tablefoot{The full sample, the dustless, and the not-dustless samples are defined in Sect. \ref{sec:colour_distribution}. The dustless sample comprises SNe Ia that are located in the outskirts of their host galaxies and in locally low-mass environments.}
\end{table}

To further investigate our physical interpretation of the model, we split our sample according to various environmental properties. If they are indeed intrinsic, the Gaussian parameters $c_\mathrm{int}$ and $\sigma_c$ should be independent of the environment, while the dust-related parameter $\tau$ should become insignificant in dust-free environments.
We chose five parameters to characterise the SN~Ia environment that are all described in Sect. \ref{sec:ztfdr2}: colour $(g-z)$ (local and global), stellar mass (local and global), and the dimensionless directional light ratio \citep[dDLR, a normalised measure of the distance between a SN and a neighbouring galaxy, see][]{Sullivan_2006, Gupta_2016}. 
We then fitted the subsamples in two ways:
\begin{enumerate}
    \item We fitted the full model on each subsample. 
    This allowed us to assess whether  $c_\mathrm{int}$ and $\sigma_c$ varied as a function of the environment.
    \item We fitted only $\tau$ per subsample but fixed the supposedly intrinsic distribution parameters ($c_\mathrm{int}$, $\sigma_c$) to the values of the full sample (as was done in \citealt{Brout_Scolnic_2021}).
\end{enumerate}

\begin{figure*}
    \centering
    \includegraphics[width=1.8\columnwidth]{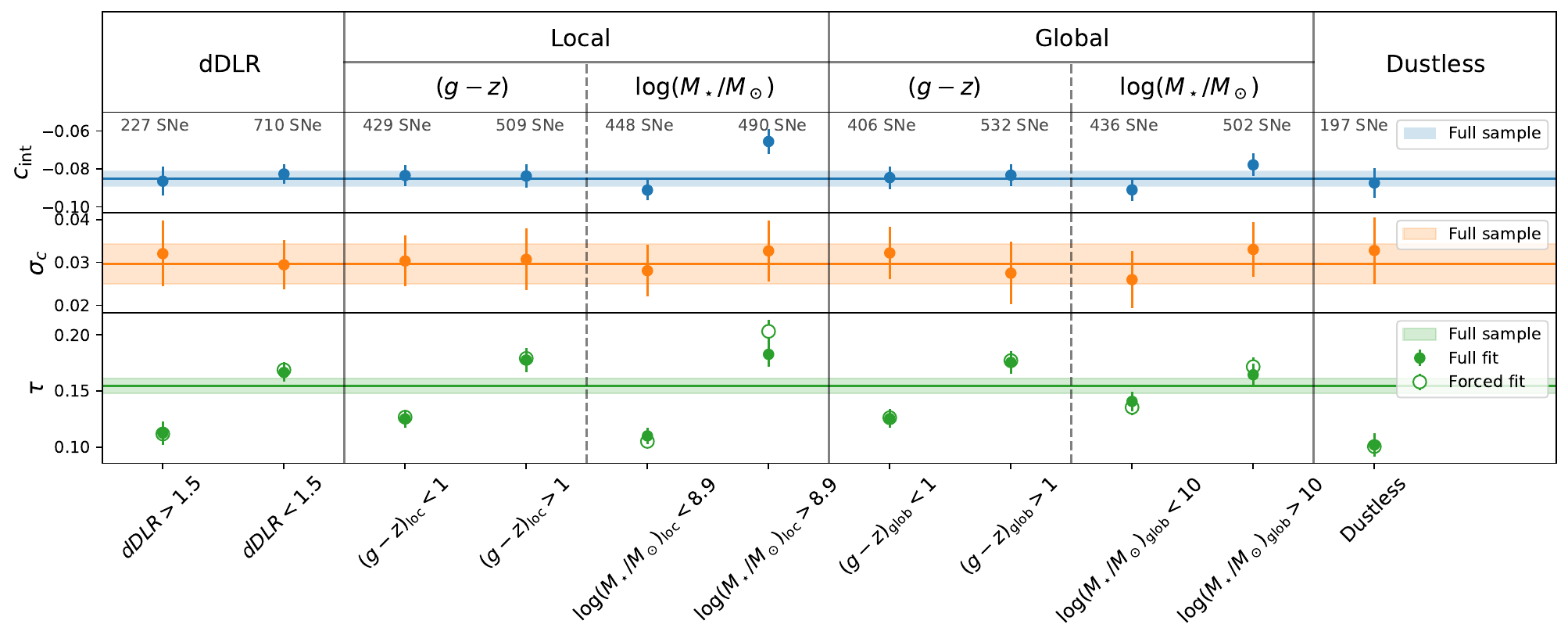}
    \caption{Parameters for the colour distribution fit described in Eq. \ref{eq:dust} for the full sample (shaded bands) and for the for subsamples split into dDLR, environmental colour, and stellar mass (global and local), and the dustless sample. The number of SNe in each subsample is indicated above the corresponding points. The full points in the bottom plot show the refitted full model ($c_\mathrm{int}$, $\sigma_c$, and $\tau_\mathrm{dust}$) on the subsample, and the empty points show the fixed intrinsic colour distribution with the fitted dust distribution.}
    \label{fig:colour_comparison}
\end{figure*}

The resulting fitted parameters are shown in Fig.~\ref{fig:colour_comparison}.
As illustrated in this figure, $c_\mathrm{int}$ and $\sigma_c$ appear to be independent of the environment. 
The sole visible variation is the redder average colour $c_\mathrm{int}$ of SNe~Ia from locally massive environments ($2.6\sigma$). When we account for the look-elsewhere effect (LEF), however, this variation is only at the $2.0\sigma$ level. 
The LEF accounts for the fact that the odds ($p_{\mathrm{LEF}}$) of finding a low-probability event ($p_{\mathrm{obs}}$) increase with the number of trials ($p_{\mathrm{LEF}} = p_{\mathrm{obs}} \times n_{\mathrm{trials}}$). In the case of Fig. \ref{fig:colour_comparison}, $n_{\mathrm{trials}}=5$ because we computed the difference in $c_\mathrm{int}$ using five different environmental cuts. Since none of the variations in $c_\mathrm{int}$ and $\sigma_c$ in the environments are significant, we conclude that they indeed account for an intrinsic SN~Ia colour variability. While we find no significant evolution of $c_\mathrm{int}$ with environment, we note that \cite{Wang_2013, Pan_2020} reported that high-velocity SNe Ia, which are intrinsically redder \citep{Burgaz_2024a}, preferentially inhabit high-mass hosts and the inner regions of those hosts, which indicates the same direction as in Fig. \ref{fig:colour_comparison}. However, this preference is not visible in our dataset (see \citealt{Burgaz_2024b}).
Another consequence of this model is that the colour range above $c>0.2$ seems largely dominated by host reddening because the Gaussian distribution representing the intrinsic SN colour distribution is almost null outside of $-0.175<c<0.005$ ($3\sigma_c$ interval around $c_\mathrm{int}$). 
The $\tau$ parameter, however, does vary as a function of the environment. SNe~Ia in the outskirts of their host galaxy ($\mathrm{dDLR}>1.5$) or in regions with a low stellar mass display significantly lower $\tau$ values than those in the centres of their galaxy or in locally massive environments ($4.1\sigma$ and $5.4\sigma$, respectively). This is in accordance with the literature \citep[e.g. ][]{Galbany_2012}. These variations are not as pronounced for global properties or local colour.
These findings support the claim that the reddening that is parametrised by the exponential decay parameter $\tau$ is caused by interstellar dust extinction because the galaxy cores and denser environments are significantly dustier than others (see Sect. \ref{sec:discussion_dust} for a discussion of dust in galaxies). 

Investigating the SN colour distribution in inner and outer regions of host galaxies with the DES-SN5YR sample, \cite{Toy_2024} reported similar trends, but with a smaller $\tau$ difference for the split on dDLR. Additionally, the evolution of the SN Ia colour distribution as a function of the stellar mass and redshift of the host was further discussed by \cite{Popovic_2024}.

We then defined a dustless SNe~Ia sample of 197 objects that lie in regions with a locally low stellar mass and in the outskirts of their host galaxy, following the cuts in Fig.~\ref{fig:colour_comparison}. With these cuts, significant host dust extinction along the line of sight of these SNe~Ia is unlikely. In contrast, SNe~Ia that are not in this sample may be reddened by dust, but not all will be.
The colour distribution for each subsample and the fitted model are shown in the bottom of Fig.~\ref{fig:colour_distrib}. The strong difference between the $\tau$ parameters for the dustless and not-dustless samples ($4.9\sigma$; see Table~\ref{tab:colour_model}) is clearly visible in Fig.~\ref{fig:colour_distrib}. Moreover, there is a $4.3\sigma$ difference in $\tau$ between the dustless and full samples. SNe~Ia from the dustless sample have a much shorter red tail than other SNe~Ia and an apparent mode shift that is caused by the reduced non-centred exponential decay convolution in Eq.~\ref{eq:dust}. The $c$ distribution is not purely Gaussian, however, suggesting that even though the dust reddening is significantly reduced, it still exists. This is expected because dust does exists in low-mass regions or at the edges of galaxies, but in smaller amounts.

\begin{figure}
    \centering
    \includegraphics[width=1\columnwidth]{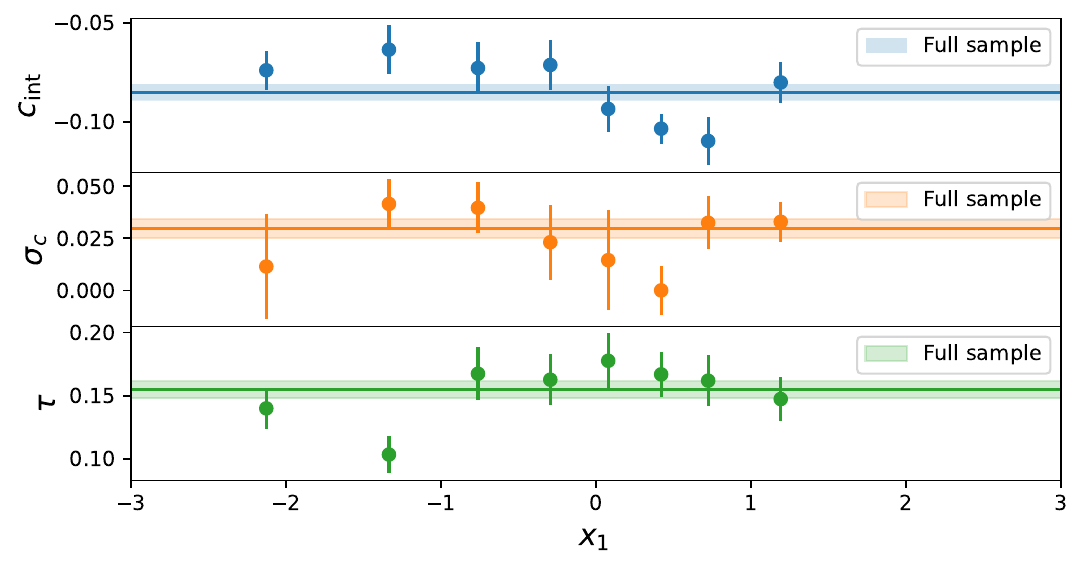}
    \caption{Parameters of the fitted colour distribution from Eq. \ref{eq:dust} $c_\mathrm{int}$, $\sigma_c$, and $\tau$ in bins of stretch. The shaded bands correspond to the full sample values, and each stretch bins contains $\mathcal{O}(120)$} SNe.
    \label{fig:colour_vs_stretch}
\end{figure}

Finally, we plot in Fig. \ref{fig:colour_vs_stretch} the evolution of the parameters from Eq. \ref{eq:dust} ($c_\mathrm{int}$, $\sigma_c$, $\tau$) against SN stretch $x_1$. By construction, the \texttt{SALT2.4} colour and stretch should be independent. Neither parameter evolves statistically significantly in Fig. \ref{fig:colour_vs_stretch}. For every stretch bin, the parameters are compatible with the full sample values at the $2\sigma$ level.

\section{Colour-residual relation}
\label{sec:colour_mag_relation}

In this section, we analyse the environmental dependence and linearity of the SN~Ia colour standardisation. This section is the colour counterpart of the stretch-standardisation study of \cite{Ginolin_2024a}, who provided a full description the standardisation procedure in their Section~4. We briefly describe the key elements below.

We computed the Hubble residuals, $\Delta\mu=\mu_\mathrm{obs}-\mu_\mathrm{cosmo}$, assuming \cite{Planck_2020} cosmology. For a volume-limited sample like ours, the observed distance moduli can be written as \citep{Tripp_1998}
\begin{equation}
    \label{eq:standardisation}
    \mu_{\text{obs}} = m_B-M_0 - \beta c + \alpha x_1 -\gamma p,
\end{equation}
with $c$ (colour), $x_1$ (stretch), and $m_B$ (peak magnitude in B band) estimated per SN using the \texttt{SALT2} T21 light-curve fitter \citep{Guy_2010, Betoule_2014,Taylor_2021}. The $p$ term is the probability for a given SN~Ia to belong to a given environmental subgroup (e.g. SNe in locally blue environments), and it takes the errors on the environmental proxy into account.

The standardisation parameters $\alpha$, $\beta$, and $\gamma$ represent the empirical standardisation relations for stretch, colour, and environment, respectively. $M_0$ is the absolute SNe~Ia magnitude, which is degenerate with the Hubble constant. We focused on the $\beta$ standardisation term and on its connection with the environmental step ($\gamma$). The stretch standardisation parameter ($\alpha$) and the environmental magnitude offset $\gamma$ were studied in detail in \cite{Ginolin_2024a}.

We fitted the standardisation parameters along with an intrinsic scatter ($\sigma_\mathrm{int}$) using a total-$\chi^2$ fit. As detailed in \cite{Ginolin_2024a}, this technique, which simultaneously fits for the true underlying $x1$, $c$, and $p$, outputs unbiased standardisation parameters. This is different from the simple $\chi^2$ method, which assumes that the measured (and noisy) $x1$, $c$, and $p$ correspond to the truth. This is demonstrated in Section 4 and in the appendix of \cite{Ginolin_2024a} and is further illustrated in Appendix~\ref{ap:fitting} for the $\beta$ standardisation.

As reported in \cite{Ginolin_2024a}, we find $\beta=3.05\pm0.06$, $\alpha=0.161\pm0.010$, $\gamma=0.143\pm0.025$, using local colour as an environmental tracer for the same sample as we used in this paper.
\cite{Ginolin_2024a} reported that the magnitude-stretch relation is significantly non-linear ($>10\sigma$), with a breaking point near $x_1\sim-0.5$. To account for this non-linearity, we replaced the $\alpha$ term in Eq.~\ref{eq:standardisation} by $\mathcal{A}(x_1)$, such that
\begin{align}
    \label{eq:brokerstandardisation_alpha}
    \mathcal{A}(x_1) =
    \begin{cases}
    \alpha_\mathrm{low}  &\text{if $x_1 < x_1^0$}\\
    \alpha_\mathrm{high} &\text{if $x_1 \geqslant x_1^0$}.
    \end{cases}
\end{align}
When we account for this non-linearity, we find $\beta=3.31 \pm 0.03$, $\alpha_\mathrm{low}=0.271 \pm 0.011$, $\alpha_\mathrm{high}=0.083 \pm 0.009$, and $\gamma=0.175 \pm 0.010$, with $x_1^0=-0.48\pm0.08$. Because this non-linearity is significant, it appears legitimate to study the colour-magnitude relation using a non-linear magnitude-stretch relation. For the sake of comparison with the literature, however, we also performed our analysis using the usual linear $\alpha$ term. The linearity of the colour-magnitude relation is studied in Sect.~\ref{sec:beta_linearity}, and its environmental dependence is examined in Sect. \ref{sec:beta_env}.

\subsection{Linearity of the colour-residual relation}
\label{sec:beta_linearity}

\begin{figure}
   \centering
   \includegraphics[width=1\columnwidth]{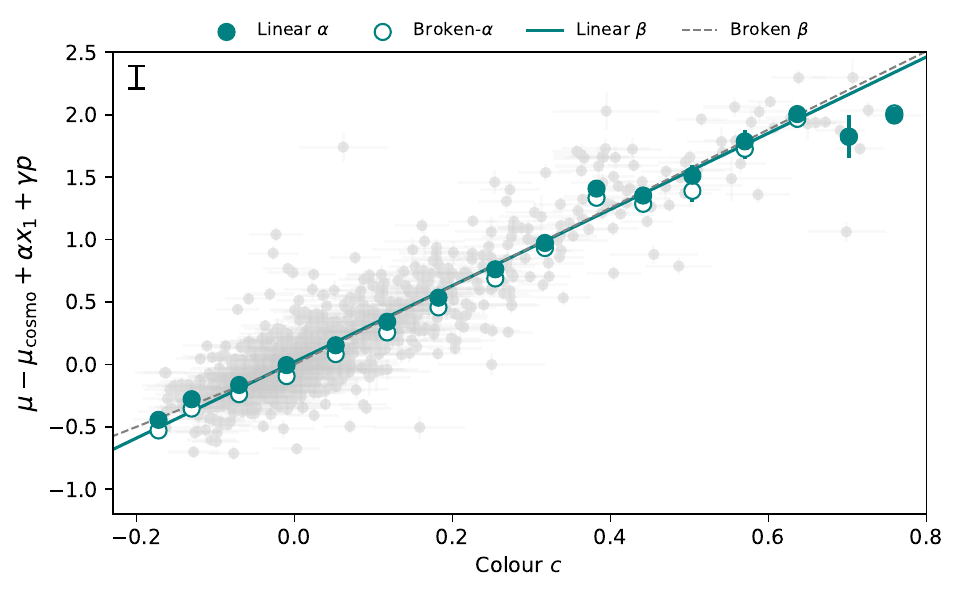}
      \caption{Binned Hubble residuals (corrected for stretch and environment) vs. colour. The error bars on the points only account for the error on the mean, but a visual indication of the fitted intrinsic scatter is given in the top left corner. The full line is the usual Tripp relation, and the dashed line is the fitted broken-$\beta$ standardisation. The open points represent the Hubble residuals corrected for stretch using the broken-$\alpha$ standardisation found in \cite{Ginolin_2024a}.}
         \label{fig:broken_beta}
\end{figure}

We plot in Fig.~\ref{fig:broken_beta} the relation between SN colour and Hubble residuals not corrected for colour to investigate the linearity of the colour-residual standardisation relation. This relation appears to be linear in the full colour range $c \in [-0.2, 0.8]$. While consistent with earlier findings \citep[e.g.][]{rose2022}, this linearity is still surprising given the discussion of the duality of the colour origin discussed in Sect.~\ref{sec:colour_distribution}.
In that section, we showed that bluer SN~Ia colours are drawn from an  environment-independent Gaussian distribution, while redder colours are likely due to additional interstellar dust reddening, which dominates at $c>0.2$. Yet, while the physical origin of the colour seems to differ, we note no inflection at high $c$ in Fig.~\ref{fig:broken_beta}, as might be expected if dust and the origin of the intrinsic colour scatter (e.g. temperature) had different colour-magnitude relations. When we fitted for two $\beta$ for two subpopulations split at a fixed cut $c^\mathrm{cut}=0$, no statistically significant deviation appeared ($\Delta \beta = -0.71\pm0.26$, a $2.7\sigma$ significance). This result agrees with the literature \citep[e.g.][]{Sullivan_2010, Rigault_2020, rose2022} and is reassuring for the use of SNe~Ia as accurate cosmological distance indicators.

As an additional test, we also fitted for three and four $\beta$. We used bins of equal sizes, comprised of $\sim 300$ and $\sim200$ SNe~Ia each. The strongest deviation from the full sample value is a lower $\beta$ in the bluest bin ($\leq2.5\sigma$) for both cases ($\beta=2.14\pm0.36$ and $\beta=1.78\pm0.50$ for the three and four $\beta$ analysis, respectively), while the other bins are all within $1\sigma$. We thus conclude that there is no significant variation of $\beta$ with SN colour.
It is still interesting, however, that it follows the expected trend from the literature. The $\beta$ value for blue SNe, which should not be affected by dust, can be compared to the value of the intrinsic $\beta_\mathrm{int}$ in analyses that explicitly modelled the dust, for instance $\beta_\mathrm{int}=1.98\pm0.18$ in \cite{Brout_Scolnic_2021} and $\beta_\mathrm{int}=2.064\pm0.174$ in \cite{Popovic_2023}.

Finally, with $\beta=3.05\pm0.06$ ($3.31\pm0.03$ with a broken-$\alpha$ standardisation), which corresponds to an $R_V\sim2.3$, we confirm that the SN~Ia colour law is significantly lower than expected for Milky Way-like dust \citep[$R_V\sim3.1$, see e.g.][]{Goobar_2008} if the colour-residual relation were fully due to dust extinction. As illustrated in Fig.~\ref{fig:broken_beta}, our conclusions are not affected when we account for the non-linearity of the stretch-magnitude relation (broken-$\alpha$).

\subsection{$\beta$ dependence on the environment}
\label{sec:beta_env}

\begin{figure}
   \centering
   \includegraphics[width=1\columnwidth]{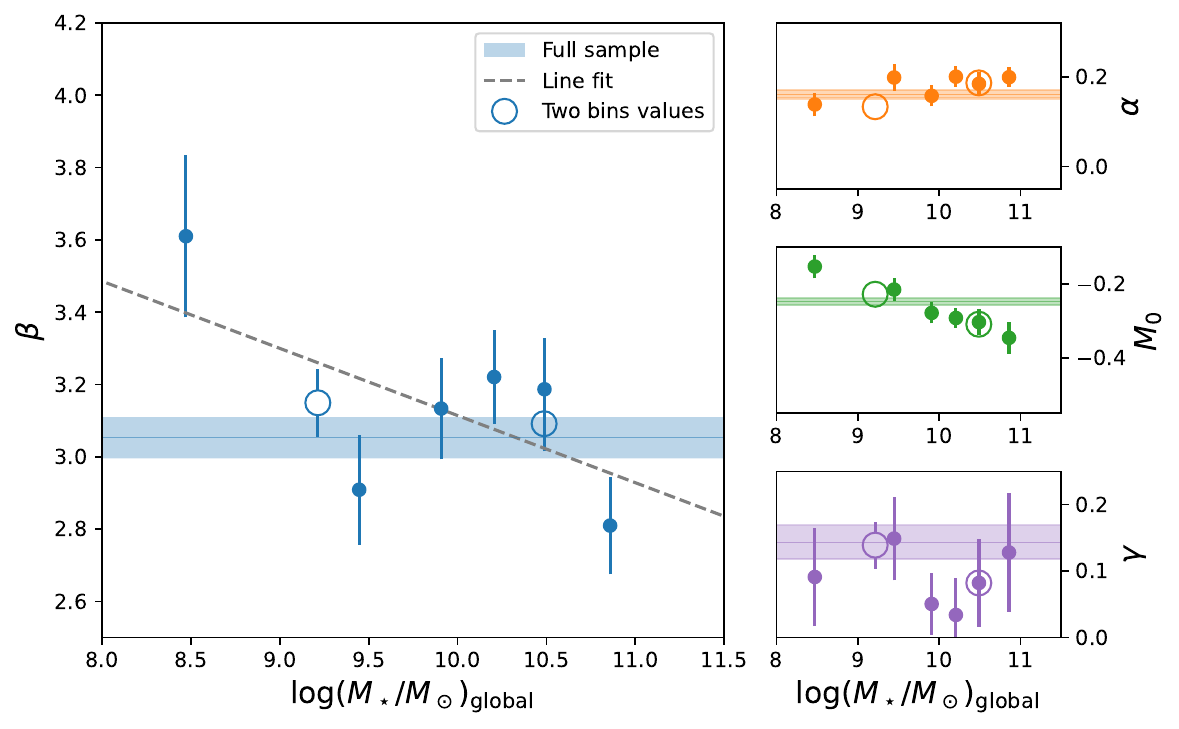}
      \caption{$\beta$, $\alpha$, $M_0$, and the local colour step $\gamma$ for each of the global mass bins. The shaded band shows the $1\sigma$ error around the fitted value for the full sample. The open points are values for when the sample is cut into two.}
         \label{fig:beta_evol}
\end{figure}

\begin{figure}
   \centering
   \includegraphics[width=1\columnwidth]{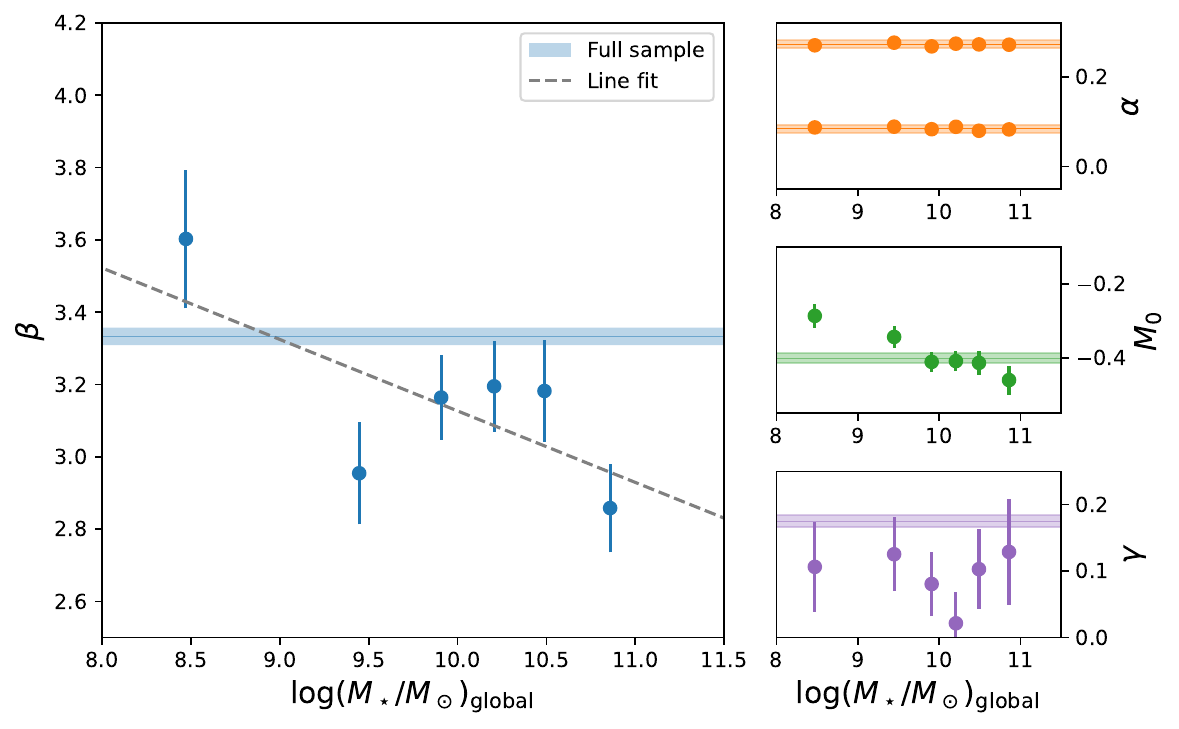}
      \caption{Same as Fig. \ref{fig:beta_evol}, but for a broken-$\alpha$ standardisation \citep{Ginolin_2024a}. A strong prior is imposed on $\alpha_\mathrm{low}$ and $ \alpha_\mathrm{high}$ to force them to stay within their error bars from the full sample.}
         \label{fig:beta_evol_broken}
\end{figure}

\begin{figure}
   \centering
   \includegraphics[width=1\columnwidth]{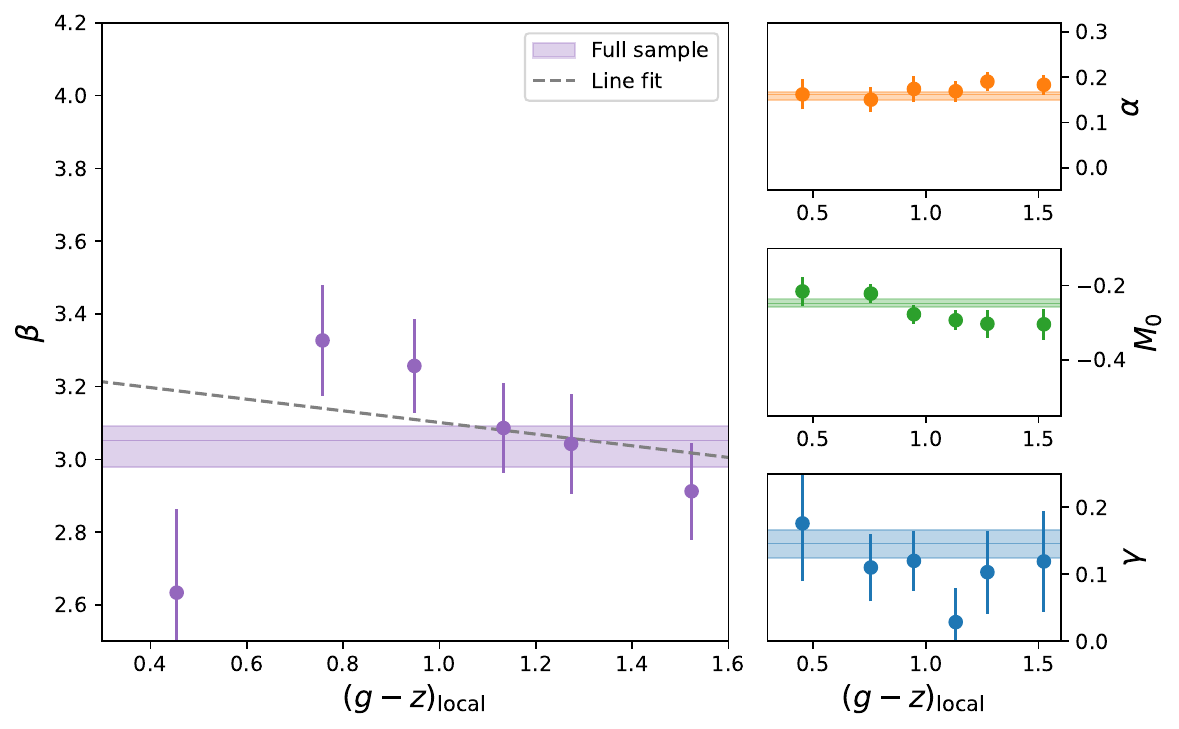}
      \caption{Same as Fig. \ref{fig:beta_evol}, but for bins of local colour and a global mass step.}
         \label{fig:beta_evol_color}
\end{figure}

We further investigated the universality of the colour-magnitude relation by studying the variation in $\beta$ as a function of the environment.
We split our sample into six equally populated bins of global stellar mass containing $\mathcal{O}(150)$ SNe~Ia each. 
We then independently standardised each of these bins, that is, we fitted for a set of ($\alpha$ or ($\alpha_\mathrm{low}$, $\alpha_\mathrm{high}$), $\beta$, $M_0$, $\gamma$) for each global mass bin.
The resulting standardisation parameters are shown in  Fig.~\ref{fig:beta_evol} for the linear stretch standardisation and in Fig.~\ref{fig:beta_evol_broken} for the broken-$\alpha$ following \cite{Ginolin_2024a}.
When applying the broken stretch-magnitude relation, we fixed the $\alpha_\mathrm{low}$ and $\alpha_\mathrm{high}$ at the values fitted on the full sample because low-mass galaxies host no low-stretch SNe~Ia (see \citealt{Ginolin_2024a}, or, e.g., \citealt{Nicolas_2021}). To do this, we implemented a Gaussian prior centred on the best-fit values, with a width corresponding to the fitted errors.

In Figs. \ref{fig:beta_evol}-\ref{fig:beta_evol_broken}, $\beta$ seems to vary as a function of stellar mass of the host. With $\beta=3.61\pm 0.22$ (linear-$\alpha$), the lowest bin ($6.5<\log(M_*/M_\odot)<9.2$) significantly favours higher $\beta$ values than higher-mass hosts ($\beta=3.05$ on average for $\log(M_*/M_\odot)\geq 9.2$). 
Fitting an affine function on the $\beta$-$\log(M_*/M_\odot)$ relation in Fig. \ref{fig:beta_evol}, we find $\beta=(-0.19\pm0.09) \times \log(M_\star/M_\odot)+(5.0\pm1.0)$, with the slope being $2.0\sigma$ away from zero. Computing the difference in the Akaike information criterion \citep[AIC, ][]{AIC} between this affine function and a constant, we find $\Delta$AIC=1.9, which means that an evolving $\beta$ is slightly favoured over a constant one. When accounting for the $\alpha$ non-linearity, we find $\beta=(-0.20\pm0.08) \times \log(M_\star/M_\odot)+(5.1\pm0.8)$, the slope being $2.4\sigma$ away from 0. In this case, an evolving $\beta$ is favoured over a constant with $\Delta$AIC=3.6.
Fig.~\ref{fig:beta_evol} and Fig.~\ref{fig:beta_evol_broken} show, however, that the $\beta$-$\log(M_\star/M_\odot)$ relation does not look linear, and only the lower-mass bin is significantly greater than the other bins. In the $\log(M_\star/M_\odot)>9.2$ range, the measured $\beta$ are consistent with a random fluctuation around the average  value.

The evolution of the other standardisation parameter is also interesting. For the linear stretch standardisation in Fig.~\ref{fig:beta_evol}, we note that $\alpha$ slightly shifts towards higher values for higher-mass hosts. This is a direct consequence of the non-linearity of the stretch-magnitude relation presented in \cite{Ginolin_2024a}, and was discussed in their Section~4. Since high-mass hosts favour low-stretch SNe~Ia (e.g \citealt{Ginolin_2024a, Rigault_2020,Nicolas_2021}) and since these have a stronger stretch-magnitude relation than higher-stretch SNe~Ia, a unique $\alpha$ leads to an increase in the apparent linear coefficient. A detailed modelling of this effect is ongoing and will be the subject of a future publication.

We now focus on the evolution of the offset term $M_0$. Although $\beta$ was left free and we accounted for a colour step per mass bin, the offset term significantly varied with host mass. This implies that the mass step is not absorbed by the other terms, as suggested for instance by \cite{Roman_2018,Rigault_2020, Kelsey_2023}. Interestingly, the $M_0$-$\log(M_\star/M_\odot)$ relation does appear to be linear in Fig.~\ref{fig:beta_evol} and Fig.~\ref{fig:beta_evol_broken}, with $M_0=(-0.012\pm0.018)\times\log(M_\star/M_\odot)+(0.54\pm0.18)$. Finally, when the non-linearity of the stretch-magnitude relation is accounted for (which was clearly demonstrated ($>14\sigma$) in \citealt{Ginolin_2024a}), the colour step seems to be stable and independent of the host mass near $\gamma\sim0.1$, even when we allowed a change in the magnitude offset per bin of mass, which absorbs part of the environmental dependences.

We show a different perspective of this analysis in Fig.~\ref{fig:beta_evol_color}, where we invert the roles played by local colour and host mass. We split the sample into six equally populated bins of local environmental colour (2~kpc radius $(g-z)$) while fitting for a mass step per bin. In this case, $\beta$ does not vary significantly, with $\beta=(-0.16\pm0.20) \times (g-z)+(3.26\pm0.22)$, and the slope is only $0.8\sigma$ away from zero.

We thus conclude from this detailed study of the standardisation parameters as a function of SN~Ia environment that (1) $\beta$ depends on the stellar mass of the host, such that lower-mass hosts ($\log(M_\star/M_\odot)<9.2$) exhibit a higher $\beta$ than other targets. This low-mass host $\beta$ value is compatible with the expectation from dust models ($\beta\sim4.1$) if the colour-magnitude relation is indeed driven by Milky Way-like dust. (2) Accounting for $\beta$ variations does not cancel the mass step and/or the local colour step, which seems to play a significant role in SN~Ia standardisation. The accurate modelling of SN Ia astrophysical dependences thus seems to be more complicated than expected.

A different $\beta$ for high- or low-mass host SNe was previously reported in \cite{Sullivan_2010, Rubin_2015, Rubin_2023}. \cite{Gonzalez-Gaitan_2021} reported a $2.9$ difference in $\beta$ when fitting in two $\beta$ bins, and no reduction of the mass step when fitting multiple (up to 4) $\beta$, in accordance with our results. In a companion paper, \cite{Popovic_2024} presented a similar analysis including high-redshift data. They drew similar conclusions concerning the $\beta$-host mass relation, which does not seem to vary as a function of redshift.
In the next section, we further investigate the \cite{Brout_Scolnic_2021} model, which suggests that $\beta$ evolves as a function of host mass and claims that this is the root cause of the other astrophysical biases \citep[see also][]{Popovic_2023, Wiseman_2023}.

\section{Colour dependence of SN Ia standardised magnitudes (steps)}
\label{sec:steps}

In this section, we investigate the dependence of the environmental step on SN colour.  \cite{Brout_Scolnic_2021} modelled the origin of the residual SN~Ia magnitude environmental step as variations in the dust content (related to $c$) and properties (related to $\beta$) as a function of the stellar mass of the host. 
In Sect.~\ref{sec:colour_distribution}, we confirmed that the colour distribution indeed seems to be twofold, and that redder SNe Ia obtain their colour excess from host-dependent properties and likely from interstellar dust. 
To test the hypothesis of a dust origin of the steps, steps for different colour bins can be computed. If dust is the main driver of the observed magnitude step, bluer SNe (i.e. not reddened by dust) should not exhibit a step. For red SNe, a step should appear. This would signal two different colour-correcting laws (i.e. two $\beta$) for the two different environments. The hypothesis of the existence of different $\beta$ for different host environments has been challenged in Sect. \ref{sec:beta_linearity}, but not as a function of SN colour.

\begin{table}
\centering
\small
\caption{Environmental steps for blue ($c<0$) and red ($c>0$) SNe.}
\label{tab:steps_blue_red}
\begin{tabular}{l c c  c } 
\hline\\[-0.8em]
\hline\\[-0.5em]
$\alpha$ & Tracer & Blue steps & Red steps \\
\hline\\[-0.5em]
Linear& $(g-z)_\mathrm{local}$ & $0.136\pm0.037$ & $0.141\pm0.028$ \\[0.30em]

 & $\log(M_\star/M_\odot)_\mathrm{global}$ & $0.134\pm0.031$ & $0.149\pm0.027$  \\[0.30em]
\hline
Broken & $(g-z)_\mathrm{local}$ & $0.151\pm0.014$ & $0.173\pm0.011$ \\[0.30em]

 & $\log(M_\star/M_\odot)_\mathrm{global}$ & $0.147\pm0.014$ & $0.162\pm0.012$  \\[0.30em]
\end{tabular}
\tablefoot{ We show steps for a linear $\alpha$ and a broken $\alpha$ standardisation (see Sect. \ref{sec:colour_mag_relation}).}
\end{table}

A first simple test for the dust hypothesis for the step origin is to cut our sample into two on SN colour, and to compute the steps separately. The values of blue ($c<0$, 379 SNe) and red ($c>0$, 559 SNe) steps are compiled in Table \ref{tab:steps_blue_red}. All of the blue SN steps are significantly different from zero. The red SNe mass steps are larger than the blue steps, but they are still consistent with the blue steps. The blue local colour steps are slightly larger than the red local colour steps, however. This indicates that dust does not account for the whole environmental step.

\begin{figure*}
   \centering
   \includegraphics[width=2\columnwidth]{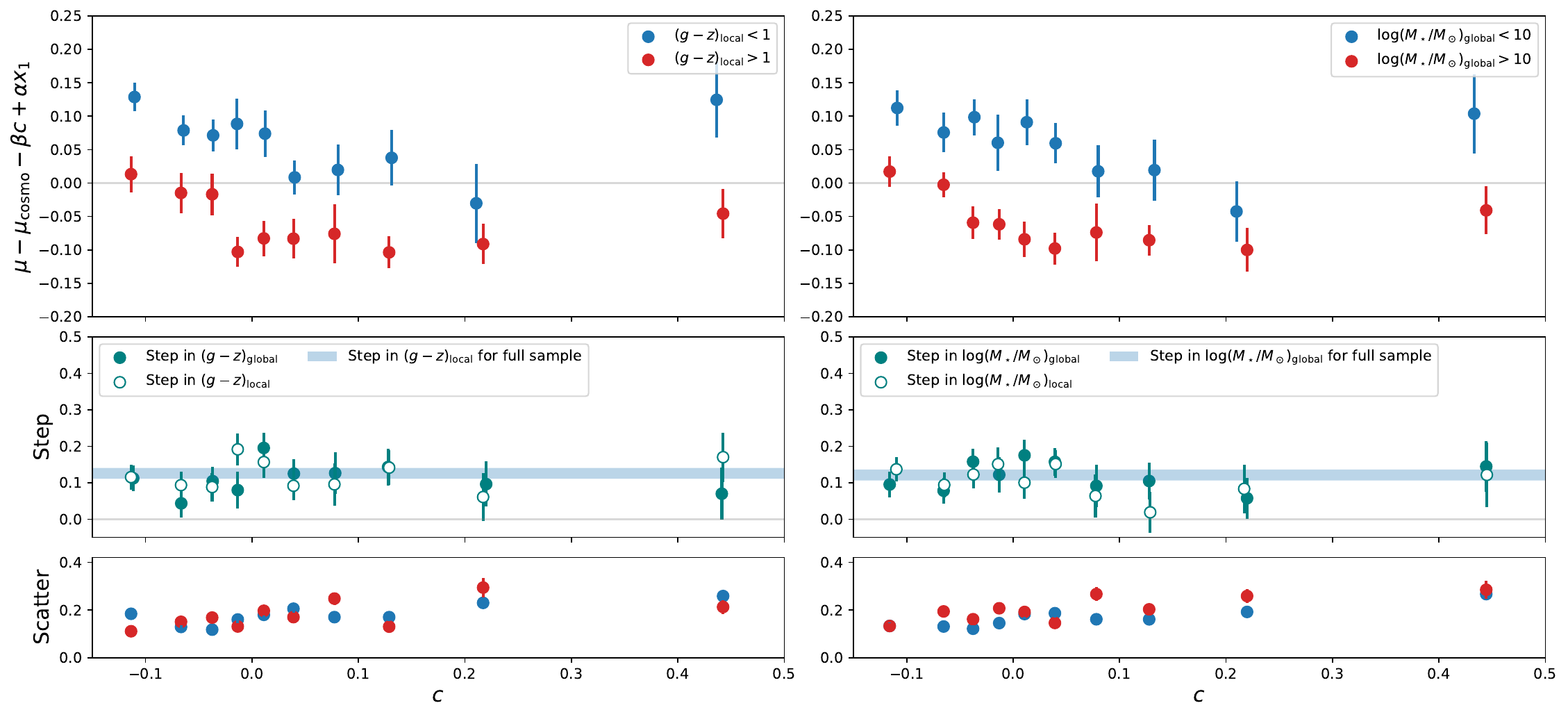}
      \caption{Same as Fig.~6 of \cite{Brout_Scolnic_2021}. \textit{Top:} Binned Hubble residuals vs. SNe colour, split according to local colour (left) and global mass (right). There are $\sim 80$ SNe in each bin. \textit{Middle:} Steps computed for each colour bin, with a split according to global and local colour (left) or mass (right). \textit{Bottom:} Scatter of the Hubble residuals for each colour bins for locally blue or red (left) or globally low- or high-mass (right) environments.}
    \label{fig:BS20}
\end{figure*}

We replicated Fig.~6 of \cite{Brout_Scolnic_2021}, as was also done in \cite{Kelsey_2023, Popovic_2023, Wiseman_2023}. We thus plot in Fig. \ref{fig:BS20} the corrected Hubble residuals, magnitude step, and scatter split on environment as a function of SN colour. In contrast to \cite{Brout_Scolnic_2021}, but as was done in \cite{Kelsey_2023}, we considered local and global colour $(g-z)$ as well as the local stellar mass as environment tracers, in addition to the global stellar mass.
The top plots in Fig. \ref{fig:BS20} clearly show that locally blue environments or low-mass host SNe~Ia are significantly fainter than locally red environments or high-mass host SNe Ia, as expected given the measured local colour and global mass steps. We highlight that the upward tilt of the Hubble residuals  at low $c$ is a direct consequence of the errors on $c$ and is not a sign of a colour-dependent $\beta$. This is explained in more detail in appendix \ref{ap:fitting}.
However, unlike \cite{Brout_Scolnic_2021}, we observe no SN colour dependence in the shape of the residuals. The step amplitudes and scatter are also consistent in all colour bins.

\begin{table}
\centering
\small
\caption{Parameters of the step evolution with SN colour.}
\label{tab:step_evolution}
\begin{tabular}{l  c  c } 
\hline\\[-0.8em]
\hline\\[-0.5em]
Tracer & Slope & $\Delta$AIC\\
\hline\\[-0.5em]
$(g-z)_\mathrm{local}$ & $0.05\pm0.11$ (0.5$\sigma$) & 1.8 \\[0.30em]
$(g-z)_\mathrm{global}$ & $0.02\pm0.12$ (0.2$\sigma$) & 2.0 \\[0.30em]
$\log(M_\star/M_\odot)_\mathrm{local}$ & $-0.11\pm0.12$ (0.9$\sigma$) & 1.2 \\[0.30em]
$\log(M_\star/M_\odot)_\mathrm{global}$ & $0.02\pm0.11$ (0.1$\sigma$) & 2.0  \\[0.30em]
\hline
\end{tabular}
\tablefoot{Slope of the fit and $\Delta$AIC between an affine function and a constant.}
\end{table}

To quantify the deviation of the steps from a constant (middle plot of Fig. \ref{fig:BS20}), we fitted for both a constant and an affine function on the step-SN colour relation. The values for the slope of the fit and the $\Delta$AIC between a constant and an affine fit for each of the tracers are presented in Table \ref{tab:step_evolution}. All slopes of the step-colour relation are smaller than $1\sigma$ away from zero, and for all tracers, the difference in the AIC between a constant and an affine function is smaller than 2. This does not support an evolution of the step with colour not.
These conclusions stand when using Hubble residuals standardised with the broken-$\alpha$ from \cite{Ginolin_2024a}. The results also remain the same for fewer bins.

Finally, we plot the scatter (normalised median absolute deviation, nmad) of the residuals for SNe in different environments in the bottom plots of Fig. \ref{fig:BS20}. The scatter for redder SNe increases slightly, which might indicate a higher dust content, as dust not only dims but also scatters SN magnitudes. However, there is no difference in the scatter between SNe in different environments, as was seen in \cite{Brout_Scolnic_2021} for SNe in high- or low-mass hosts. This was not replicated by any of the following analyses either \citep{Wiseman_2022, Kelsey_2023}.

\section{Discussion}
\label{sec:discussion}

\subsection{Tests}
\label{sec:tests}

\begin{table}
\caption{Robustness tests as described in Sect. \ref{sec:tests}.}
\centering
\small
\begin{tabular}{l c c c c} 
\hline\\[-0.8em]
\hline\\[-0.5em]
Test & Sample Size & $\Delta\tau$ & $\Delta \mathrm{AIC}_{\beta~\mathrm{evol}}$ & Slope\\
     & & [$\sigma$] & & [$\sigma$]\\
\hline\\[-0.5em]
Main analysis & 945 & 4.9 & 1.9 & 2.0 \\
\hline\\[-0.5em]
$z_\mathrm{max}=0.05 $ & 613 & 4.5 & 3.3 & 2.3 \\[0.30em]
$z_\mathrm{max}=0.07 $ & 1341 & 6.1 & 7.3 & 3.1 \\[0.30em]
Host $z$ & 707 & 4.1 & 0.3 & 1.3 \\[0.30em]
incl. Ia-pec & 965 & 4.6 & 0.8 & 1.1 \\[0.30em]
discard 91t & 878 & 4.5 & 3.6 & 2.4\\[0.30em]
$c<0.3$ & 852 & 1.3 & 5.4 & 2.7 \\[0.30em]
$\chi^2_{\texttt{SALT}}<0.1$ & 735 & 4.3 & 11.5 & 3.7 \\[0.30em]
\hline
\end{tabular}
\vspace{5pt}
\tablefoot{Sample size, difference in $\tau_\mathrm{dust}$ between the dustless sample and the not-dustless sample, $\Delta$AIC between a constant and an affine function fitted on the $\beta$ evolution with mass, and significance of the slope of this affine function for the different test samples described in Sect. \ref{sec:tests}.}
\label{tab:tests}
\end{table}

To assess the robustness of our results, we tested the impact of the choices made in Sect. \ref{sec:data} on our results, namely the difference in the dust content $\tau$ between the the dustless sample and the not-dustless sample. This corresponds to the different $\tau$ parameters between the green and orange distributions in Fig. \ref{fig:colour_distrib}, and the significance of the $\beta$ evolution (here the $\Delta$AIC between a constant fit and an affine function fit of the $\beta$-global mass relation as well as the significance of the slope in $\sigma$, as done in Sec. \ref{sec:beta_env}). The tests on the $\beta$ evolution were made using a linear correction in stretch $\alpha$.
The different tests consisted of the following seven runs:
\begin{itemize}
    \item Lowering $z_\mathrm{max}$. This strengthens the volume-limited claims.
    \item Increasing $z_\mathrm{max}$. This increases the statistics with a limited bias in the sample.
    \item Only using redshifts from host spectra. This tests the impact of the redshift errors on the sample.
    \item Including peculiar SNe Ia. This mimics the contamination that might occur if a photometric classification were used alone. 
    \item Discarding 91t. This confirms that our results do not depend on this subpopulation.
    \item Using a stronger cut on colour ($c<0.3$). This checks that our results are not reliant on redder objects.
    \item Enforcing a stronger cut on the ligth-curve fit quality. This ensures that our results are not due to poorly fitted objects.
\end{itemize}
For the $\beta$ evolution analysis, the bins were fixed for all the tests for a fair comparison.

The results for each test are presented in Table \ref{tab:tests}.
The significance of our results does not differ strongly for $z_\mathrm{max}=0.05$. The statistical boost of assuming $z_\mathrm{max}=0.07$ increases the significance of $\Delta\tau_\mathrm{dust}$ as well as the $\beta$ evolution. The use of host redshifts alone drastically lowers the $\beta$ evolution significance. This is likely due to the different host mass distributions described in \cite{Ginolin_2024a}. The host-$z$ sample indeed probes higher masses than the full sample because it is easier to obtain a galaxy spectrum for massive galaxies. Thus, the low-mass bin of the $\beta$ evolution analysis is underfilled, and the resulting $\beta$ has a very high error bar, while its value is consistent with the full sample. The inclusion of peculiar SNe Ia does not affect the dust analysis, but the significance of the $\beta$ evolution decreases. This is due to a lower $\beta$ in the lowest mass bin. This might be a statistical fluctuation, as this change only adds 20 objects to the sample. When we discarded 91t, $\Delta\tau$ or the $\beta$ evolution were not affected. When we only used $c<0.3$ SNe, the significance of $\Delta\tau$ was strongly reduced. This is expected because $\tau$ is mostly constrained by the red tail of the colour distribution. However, the $\beta$ evolution tests are similar, in accordance with the conclusion of Sect. \ref{sec:beta_linearity} that $\beta$ is the same for the full colour range. A stricter cut on $\chi^2_\mathrm{SALT}$ produces a stronger $\beta$ evolution, as the error bar on the low-mass bin is lower than for the full sample case. This might be due to smaller errors on $c$ on average because the \texttt{SALT2.4} fit is better.

\subsection{Dust in galaxies}
\label{sec:discussion_dust}

The $\tau$ parameter from Eq. \ref{eq:dust} can inform us on the dust content of the SN environments, as shown in Fig. \ref{fig:colour_comparison}.
The decreasing extinction for elliptical and spiral galaxies and for different mass ranges was studied by \cite{Goddard_2017, Gonzalez_2015, battisti2016}. As expected, the outskirts of galaxies (i.e. high dDLR regions) are less dusty than galaxy cores for early- and late-type galaxies, although the slope of the extinction-radius relation is greater for late types. Following the known stellar mass to dust mass relation \citep{Beeston_2018}, massive spiral galaxies have more dust than low-mass galaxies at a given radius. This is not visible for elliptical galaxies because the shape of the extinction-radius relation is relatively flat. Because red environments are also more massive, we expect red environments to be more dusty \citep{Lange_2015}.
These relations hold for global parameters.
For spiral galaxies, \cite{Smith_2016} showed that the distribution of dust is consistent with an exponential decay with a gradient of $\sim-1.7\,\mathrm{dex}\,R_{25}^{-1}$ \citep[with $R_{25}$ the radius where the optical brightness corresponds to a
B-band brightness of $25\,\mathrm{mag}\,\mathrm{arcsec}^{-2}$,][]{Bigiel_2012}. They further showed that this decline is very similar to that of the stellar density, which is slower than that of the star formation rate surface density. This indicates a stronger relation between the local stellar mass and dust than for local colour.
Overall, studies of the dust content in galaxies agree well with the conclusions from Fig. \ref{fig:colour_comparison}.

\section{Conclusion}
\label{sec:conclusion}

We presented an analysis of the colour term of the standardisation procedure that is used in SN Ia cosmology, of its dependence on environment, and of its link to environmental magnitude offsets (steps). A companion paper \citep{Ginolin_2024a} focused on stretch and steps. We used the volume-limited version of the ZTF DR2 SN Ia sample, which is a sample of 945 SNe Ia. This sample is free from selection effects, which allowed us to study true underlying distributions and correlations without any biases.

Our conclusions are listed below.

\begin{itemize}
    \item We probed the SN colour up to $c=0.8$. Its distribution is well characterised by the convolution of a Gaussian (thought to be the intrinsic colour distribution) and an exponential tail (thought to be host galaxy dust reddening). 
    \item We observed  a significant reduction of the red exponential tail (4.3$\sigma$) based on environmental cuts to isolate not-dusty environments (large dDLR and locally low stellar mass environments). This suggests that this term indeed accounts for additional dust reddening that dominates the SN colour at $c>0.2$.
    \item We found that a linear $\beta$ along the full SN colour range ($-0.2<c<0.8$) matches the data surprisingly well in the colour-residual relation. We found no deviations near $c\sim0.2$, for instance. This suggests that regardless of the cause of the SN colour, the colour-magnitude relation seems to be independent of colour.
    \item We found indications of an evolution of the colour-residual relation slope $\beta$ with the host galaxy stellar mass, however. The slope is only 2.0$\sigma$ away from zero, but we found that SNe in low-mass hosts ($\log(M_\star/M_\odot)<9.2$) have $\beta=3.61\pm0.02$, while the full sample has $\beta=3.05\pm0.06$. When we accounted for the broken-$\alpha$ standardisation found in \cite{Ginolin_2024a}, the slope is at the 2.4$\sigma$ level.
    \item Finally, we found no dependence of the amplitude of the environmental magnitude offset $\gamma$ as a function of SN colour, in contrast to recent results from the literature \citep[e.g.,][]{Brout_Scolnic_2021, Kelsey_2023}. This suggests that the interstellar dust of the host may not be the parameter that drives the observed magnitude offsets.
    
\end{itemize}

\begin{acknowledgements}
Based on observations obtained with the Samuel Oschin Telescope 48-inch and the 60-inch Telescope at the Palomar Observatory as part of the Zwicky Transient Facility project. ZTF is supported by the National Science Foundation under Grants No. AST-1440341 and AST-2034437 and a collaboration including current partners Caltech, IPAC, the Weizmann Institute of Science, the Oskar Klein Center at Stockholm University, the University of Maryland, Deutsches Elektronen-Synchrotron and Humboldt University, the TANGO Consortium of Taiwan, the University of Wisconsin at Milwaukee, Trinity College Dublin, Lawrence Livermore National Laboratories, IN2P3, University of Warwick, Ruhr University Bochum, Northwestern University and former partners the University of Washington, Los Alamos National Laboratories, and Lawrence Berkeley National Laboratories. Operations are conducted by COO, IPAC, and UW.
SED Machine is based upon work supported by the National Science Foundation under Grant No. 1106171. The ZTF forced-photometry service was funded under the Heising-Simons Foundation grant \#12540303 (PI: Graham). This project has received funding from the European Research Council (ERC) under the European Union's Horizon 2020 research and innovation program (grant agreement n 759194 - USNAC). This work has been supported by the Agence Nationale de la Recherche of the French government through the program ANR-21-CE31-0016-03. This work was supported by the GROWTH project \citep{Kasliwal2019} funded by the National Science Foundation under Grant No 1545949. TEMB acknowledges financial support from the Spanish Ministerio de Ciencia e Innovaci\'on (MCIN), the Agencia Estatal de Investigaci\'on (AEI) 10.13039/501100011033, and the European Union Next Generation EU/PRTR funds under the 2021 Juan de la Cierva program FJC2021-047124-I and the PID2020-115253GA-I00 HOSTFLOWS project, from Centro Superior de Investigaciones Cient\'ificas (CSIC) under the PIE project 20215AT016, and the program Unidad de Excelencia Mar\'ia de Maeztu CEX2020-001058-M. LG acknowledges financial support from AGAUR, CSIC, MCIN and AEI 10.13039/501100011033 under projects PID2020-115253GA-I00, PIE 20215AT016, CEX2020-001058-M, and 2021-SGR-01270. UB, JHT, MD, GD and KM are supported by the H2020 European Research Council grant no. 758638. This work has been supported by the research project grant “Understanding the Dynamic Universe” funded by the Knut and Alice Wallenberg Foundation under Dnr KAW 2018.0067, {\em Vetenskapsr\aa det}, the Swedish Research Council, project 2020-03444. Y.-L.K. has received funding from the Science and Technology Facilities Council [grant number ST/V000713/1]. SD acknowledges support from the Marie Curie Individual Fellowship under grant ID 890695 and a Junior Research Fellowship at Lucy Cavendish College. 
\end{acknowledgements}

\bibliographystyle{aa}
\bibliography{aanda}

\begin{appendix}

\section{Shape of the colour-residual relation}
\label{ap:fitting}

\begin{figure}
   \centering
   \includegraphics[width=1\columnwidth]{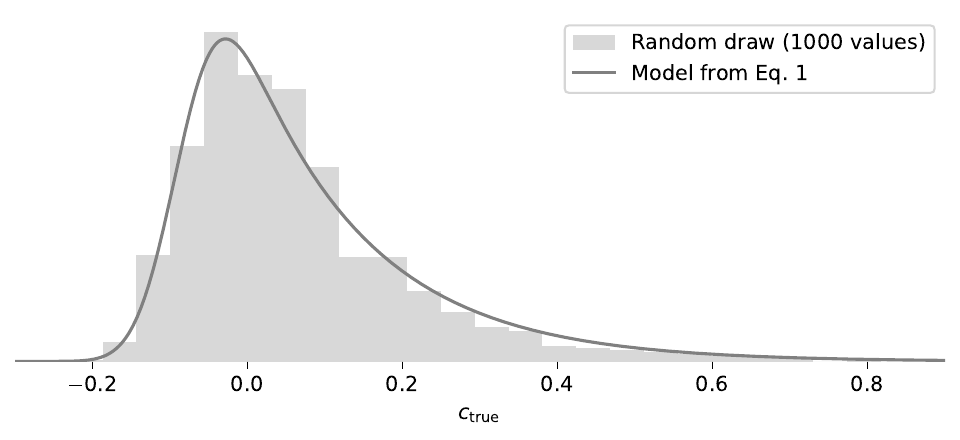}
    \caption{Histogram of the 1,000 colour values simulated using the colour model from Eq. \ref{eq:dust} with values from the full sample in Table \ref{tab:colour_model} (grey line).}
    \label{fig:simulated_colour}
\end{figure}

\begin{figure}
   \centering
   \includegraphics[width=1\columnwidth]{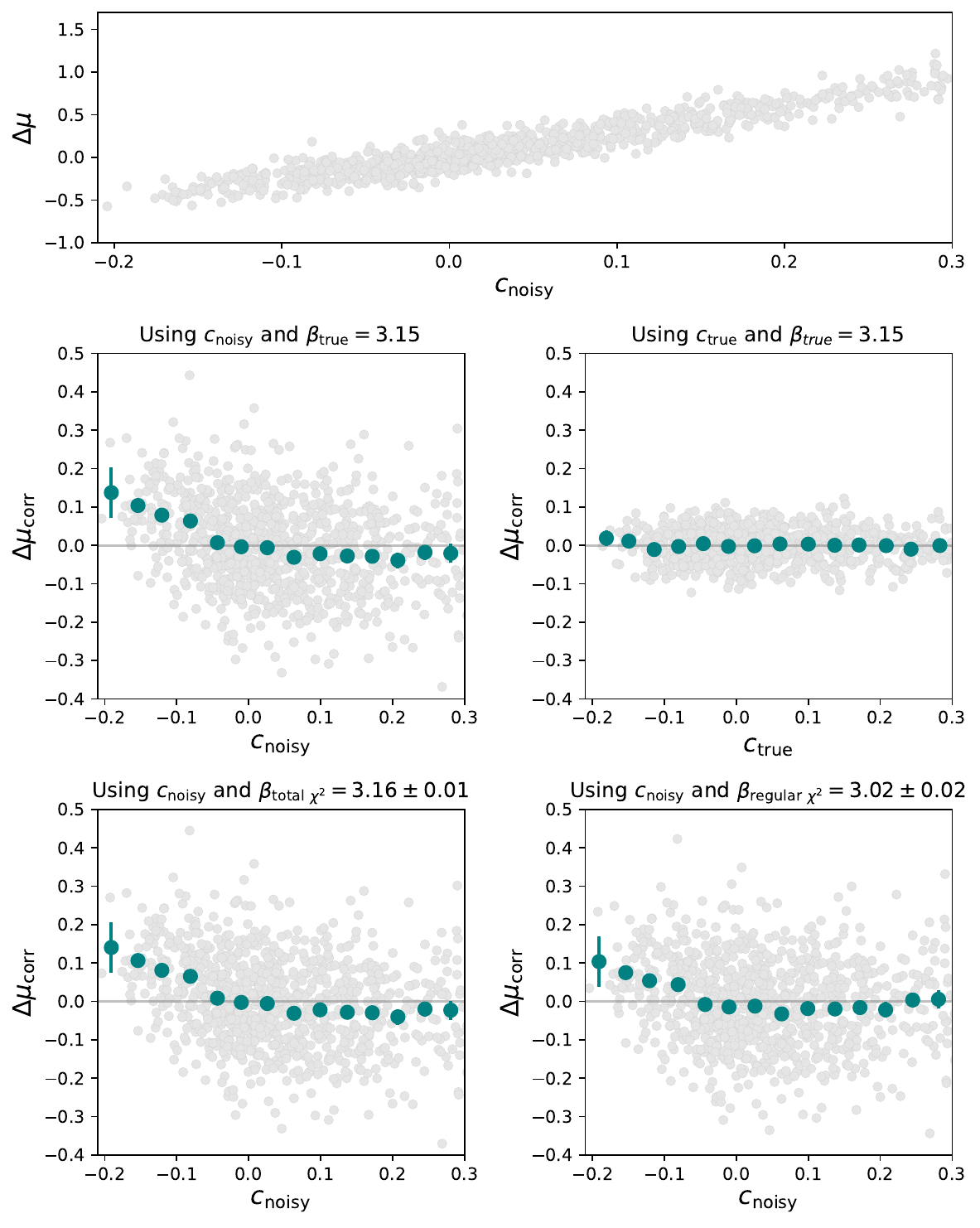}
    \caption{Simple simulation of noisy residuals with a correlation with colour $\beta$. \textit{Top:} Residuals vs. colour. \textit{Middle left:} Residuals corrected for their correlation with colour with the true correlation coefficient $\beta$ but noisy colours $c_\mathrm{noisy}$. An upwards tilt appears at low colours due to errors on $c_\mathrm{noisy}$. \textit{Middle right:} Residuals corrected for their correlation with colour with the true $\beta$ and true colours $c_\mathrm{true}$. The corrected residuals appear flat with colour. \textit{Bottom right:} Residuals corrected with an unbiased $\beta$ fitted with the total-$\chi^2$ method. \textit{Bottom left:} Residuals corrected with a biased $\beta$ fitted with the regular $\chi^2$ minimisation.}
    \label{fig:res_biais}
\end{figure}

In this appendix, we discuss the shape of the colour and stretch corrected residuals in Fig. \ref{fig:BS20}, in particular the upwards tilt of the residuals seen at lower colours. This discussion is valid for any fit of a linear relation between two noisy variables. We here only demonstrate the principle of this effect on a simple case, with a single linear relation between residuals and colour, not accounting for stretch and environment correlations, and without intrinsic scatter.
We simulated a colour sample (1,000 objects) using the model from Eq. \ref{eq:dust} and the parameters from the full sample fit presented in Table \ref{tab:colour_model}. The colour histogram along with the distribution used to generate them are plotted in Fig. \ref{fig:simulated_colour}. We then applied Gaussian noise on the true colour values $c_\mathrm{true}$ to get noisy colour values $c_\mathrm{noisy}=c_\mathrm{true}+\mathcal{N}(0, \sigma_c)$, where $\sigma_c=0.035$ is roughly the mean error on colour in our data. 
On the other hand, we computed noisy residuals, only adding a correlation with colour, so that $\Delta\mu=\beta_\mathrm{true} c_\mathrm{true}+\mathcal{N}(0, \sigma_\mu)$, with $\beta_\mathrm{true}=3.15$. $\sigma_\mu=0.04$ is the typical measurement error on magnitudes in our data. These noisy residuals are plotted against noisy colour in the top plot of Fig. \ref{fig:res_biais}.

We finally computed corrected residuals. We first used the true $\beta$ and the noisy colour to correct for the colour-residual correlation, such that $\Delta\mu_\mathrm{corr}=\Delta\mu-\beta_\mathrm{true} c_\mathrm{noisy}$. This is plotted in the middle left plot of Fig. \ref{fig:res_biais}. An upwards tilt at low colour is seen, even though we corrected with the true $\beta$.
We also computed the residuals corrected with the true colours and true $\beta$, $\Delta\mu_\mathrm{corr}=\Delta\mu-\beta_\mathrm{true} c_\mathrm{true}$, presented in the middle right plot of Fig. \ref{fig:res_biais}. In that case, the residuals appear flat with colour. We thus illustrated that the upwards tilt of the colour-residual relation seen in Fig. \ref{fig:BS20} is not the consequence of a colour-dependent $\beta$. The magnitude of this effect is stronger when errors on the $x$ variable (here $c$) are bigger.

Errors on the y-axis on the fit are the reason why a regular $\chi^2$ optimisation algorithm is biased in $\beta$ (as well as $\alpha$ and $\gamma$ when considering the full standardisation process). This is why a total-$\chi^2$ fitting method is needed in order to get unbiased standardisation parameters. The total-$\chi^2$ method fits for the "true" $x$ variables (in our case $c$) as latent parameters. 
This is illustrated in the bottom plots of Fig. \ref{fig:res_biais}. Indeed, on the bottom left plot are residuals corrected with noisy colours and $\beta$ fitted with the total-$\chi^2$ algorithm used throughout this paper. We found $\beta_{\mathrm{total~}\chi^2}=3.16\pm0.01$, close to the input value $\beta_\mathrm{true}=3.15$. On the bottom right plot of Fig. \ref{fig:res_biais} are the residuals corrected with noisy colours and $\beta$ fitted with the usual $\chi^2$ algorithm. We found $\beta_{\mathrm{regular~}\chi^2}=3.02\pm0.02$, a $6.1\sigma$ bias. The lower $\beta$ is trying to compensate for the tilt seen at low colour.
A detailed explanation of the total-$\chi^2$ fitting method can be found in Section 4.2 of \cite{Ginolin_2024a}, as well as a more thorough investigation of biases on fitted standardisation parameters using realistic ZTF sample simulations in their appendix A.

\end{appendix}

\end{document}